\documentstyle[epsfig,12pt]{article}

\newcommand{\bce}{\begin{center}} 
\newcommand{\ece}{\end{center}}
\newcommand{\beq}{\begin{equation}}
\newcommand{\eeq}{\end{equation}}
\newcommand{\bea}{\vspace{0.25cm}\begin{eqnarray}}
\newcommand{\eea}{\end{eqnarray}}

\newcommand{\brho}{\mbox{\boldmath $\rho$}}

\newcommand{\bk}{{\bf k}}

\newcommand{\br}{{\bf r}}

\newcommand{\ba}{\begin{array}}
\newcommand{\ea}{\end{array}}

\newcommand{\doublespace}{
    \renewcommand{\baselinestretch}{1.6}\large\normalsize}

\newcommand{\bb}{{\bf b}}

\def\lsim{\mathrel{\rlap{\lower4pt\hbox{\hskip1pt$\sim$}}
    \raise1pt\hbox{$<$}}}         
\def\gsim{\mathrel{\rlap{\lower4pt\hbox{\hskip1pt$\sim$}}
    \raise1pt\hbox{$>$}}}         

\def\Pom{{\bf I\!P}}

\def\beq{\begin{equation}}
\def\endeq{\end{equation}}
\def\arr{\begin{eqnarray}}
\def\endarr{\end{eqnarray}}

\makeindex

  \advance\oddsidemargin  by -1in
  \advance\evensidemargin by -1in
  \advance\topmargin      by -0.5in
\begin{document}

\vspace{2.0cm}

\begin{flushright}
\end{flushright}

\vspace{1.0cm}

\begin{center}
{\Large \bf 
Non-linear BFKL dynamics: \\
color screening vs. gluon fusion}\\

\vspace{0.5cm} 
{\large \bf R. Fiore$^{1}$, P.V. Sasorov$^2$ and  
V.R. Zoller$^2$}\\
\vspace{0.5cm}
$^1)${\em Dipartimento di Fisica,
Universit\`a     della Calabria\\
and\\
 Istituto Nazionale
di Fisica Nucleare, Gruppo collegato di Cosenza,\\
I-87036 Rende, Cosenza, Italy}

$^{2)}${\em Institute for  Theoretical and Experimental Physics,
Moscow 117218, Russia} 
\vspace{1.0cm}

{ \bf Abstract }\\
\end{center}

\vspace{1.0cm}
A feasible  mechanism of  unitarization of amplitudes of 
deep inelastic scattering at small values of Bjorken  $x$
is   the gluon fusion.
 However, its efficiency  depends crucially on  
the vacuum  color screening effect which  accompanies
 the multiplication and the diffusion of BFKL gluons from small to 
large distances. 
From the fits to lattice data 
on field strength correlators the propagation  length of 
perturbative gluons is  $R_c\simeq 0.2-0.3$ fermi. The probability to
 find a perturbative gluon with
short propagation length at large distances  is suppressed exponentially.
It changes  the pattern of (dif)fusion  dramatically. 
The magnitude  of the fusion effect appears to be 
controlled by the new dimensionless parameter $\sim R_c^2/8B$,
 with the diffraction cone slope $B$ standing for  the characteristic 
size of the interaction region.  It should slowly 
$\propto 1/\ln Q^2$ decrease at large $Q^2$. Smallness of the ratio
$R_c^2/8B$ makes the non-linear effects rather weak even at lowest
Bjorken $x$ available at HERA.
We report the results of our  studies of the non-linear BFKL equation
which has been generalized to incorporate the  running coupling and 
the screening radius $R_c$ as the infrared regulator.

\doublespace

\vskip 0.5cm \vfill $\begin{array}{ll}
\mbox{{\it email address:}} & \mbox{roberto.fiore@cs.infn.it} \\
\mbox{{\it email address:}} & \mbox{sasorov@itep.ru} \\
\mbox{{\it email address:}} & \mbox{zoller@itep.ru} \\
\end{array}$

\pagebreak


{\bf 1. Introduction.}
 
In processes of deep inelastic scattering (DIS)  the density of BFKL 
\cite{BFKL} gluons,
 ${\cal F}(x,k^2)$, 
 grows fast to smaller values of Bjorken $x$,
${\cal F}(x,k^2)\propto x^{-\Delta}$, where, phenomenologically, 
  $\Delta\approx 0.3$. 
The growth of ${\cal F}(x,k^2)$ will have to slow down
when the gluon densities become large enough that fusion processes $gg\to g$
become important. It was the original parton model  idea of 
Refs. \cite{Kancheli73, NZ75}
 developed further  within  QCD  in   \cite{GLR,MQ}.
The BFKL dynamics of saturation  of the  parton densities 
has been discussed first  in  
 \cite{B,K,Braun}, 
 for the alternative form of the fusion correction see  Eq.(A10) of 
Ref.~\cite{NSCH2006}. The literature abounds 
with suggestions of different versions of the non-linear evolution equation,
see e.g. \cite{BarKut}. 

There is, however, at least one more mechanism 
to  prevent generation of
the high density gluon states. This is well known the 
vacuum  color screening.  
The non-perturbative fluctuations in the QCD vacuum restrict the phase 
space for the 
perturbative (real and virtual) gluons introducing a new scale: 
the correlation/propagation  radius $R_c$ of 
perturbative gluons. The perturbative gluons with short propagation length,
 $R_c\sim 0.2-0.3$ fermi, as it follows from the fits to lattice data 
on field strength correlators \cite{MEGGIO}, do not walk to large distances, 
where they  supposedly  fuse together. The fusion probability decreases. 
We show that it is 
controlled by the new dimensionless parameter $R_c^2/8B$,
 with the diffraction cone slope $B$ standing for  the characteristic 
size of the region populated with interacting gluons.

The effects of finite  $R_c$ are consistently  incorporated 
 by the generalized color dipole (CD) BFKL equation 
(hereafter CD BFKL)\cite{NZZJL94,NZJETP94}.
 In   presence of a new scale the saturation phenomenon  acquires some new 
features and
the goal of  this  communication is to present their  quantitative analysis.

{\bf 2. CD BFKL  and phenomenology of DIS.}

We sketch first
the CD BFKL equation  for  $q\bar q$ dipole-nucleon  cross
section $\sigma(\xi,r)$, where $\xi=\ln(x_0/x)$ and $r$ is the $q\bar q$-separation.
 The BFKL  cross section $\sigma(\xi,r)$
sums the Leading-Log$(1/x)$ multi-gluon production cross
sections within the QCD perturbation theory (PT). Consequently, as a
realistic boundary condition for the
BFKL dynamics  one can take the lowest PT order $q\bar q$-nucleon cross section at 
some $x=x_0$. It is described by   the  Yukawa screened 
two-gluon exchange 
\bea
\sigma(0,r)\equiv\sigma_0(r)={4 C_F}\int{d^2\bk\over (k^2+\mu_G^2)^2}
\alpha_S(k^2)\alpha_S(\kappa^2)\nonumber\\
\times\left[1-J_0(kr)\right]\left[1-F_2(\bk,-\bk)\right],
\label{eq:BORN}
\eea
where
$\mu_G=1/R_c$, 
$\alpha_{S}(\kappa^2)=4\pi/\beta_0\ln(\kappa^2/\Lambda^2_{QCD})$  and 
 $\kappa^2=max\{k^2,C^2/r^2\}$. 
The two-quark form factor of the nucleon can be related to the single-quark form
factor
\beq
F_2(\bk,-\bk)=F_1(uk^2).
\label{eq:FORM2} 
\eeq
The latter is close to the charge form factor of the proton
$F_1(q^2)\approx F_p(q^2)=1/(1+q^2/\Lambda^2)^2$, 
where $\Lambda^2=0.71$ GeV$^2$ and 
in Eq.(\ref{eq:FORM2})  $u=2N_c/(N_c-1)$ for the color group  
$SU(N_c)$ \cite{Witten}.

 The small-$x$ evolution correction
to  $\sigma(\xi,r)$ for the perturbative $3$-parton state $q\bar q g$ 
is as follows \cite{NZZJL94,NZJETP94}
\bea
{\partial_{ \xi} \sigma(\xi,r)}
=\int d^{2}{\brho}_{1}\,\,
\left|\psi({\brho}_{1})-\psi({\brho}_{2})\right|^{2}\nonumber\\
\times\left[\sigma_3(\xi,\br,\brho_1,\brho_{2})-\sigma(\xi,r)\right] ,
\label{eq:CDBFKL}
\eea
where  the 3-parton ($q\bar q g$-nucleon) cross section is
\beq
\sigma_3(\xi,\br,\brho_1,\brho_2)={C_A\over 2C_F}\left[\sigma(\xi,\rho_1)+
\sigma(\xi,\rho_{2})- \sigma(\xi,r)\right]+\sigma(\xi,r),
\label{eq:SIGMA3}
\eeq
where $C_A=N_c$ and $C_F=(N_c^2-1)/2N_c$.
Denoted by $\brho_{1,2}$ are the $q$-$g$ and $\bar{q}$-$g$ separations
in the two-dimensional impact parameter plane
for dipoles generated by the $\bar{q}$-$q$
color dipole source.
 The radial light 
cone wave function $\psi({\brho})$ of the 
 dipole with the vacuum screening of infrared gluons is \cite{NZZJL94,NZJETP94}
\beq
\psi({\brho})={\sqrt{C_F\alpha_S(R_i)}\over \pi}{{\brho}\over \rho R_c}
K_{1}(\rho/R_c),
\label{eq:PSIQG}
\eeq
where $K_{\nu}(x)$ is the modified Bessel function. The one-loop  QCD coupling 
\beq
\alpha_{S}(R_i)=4\pi/\beta_0\ln(C^2/\Lambda^2_{QCD}R_i^2)
\label{eq:ALPHAS}
\eeq
 is  taken at 
the shortest relevant
distance $R_{i}={\rm min}\{r,\rho_{i}\}$. 
In the numerical analysis $C=1.5$, $\Lambda_{QCD}=0.3$ GeV,
 $\beta_0=(11N_c-2N_f)/3$  and
 infrared freezing $\alpha_{S}(r>r_f)=\alpha_f=0.8$  has been imposed 
(for more discussion see \cite{Slava2007}).
 The scaling BFKL equation \cite{BFKL} is obtained from Eq.~(\ref{eq:CDBFKL})
at $r,\rho_{1,2} \ll R_{c}$
in the  approximation $\alpha_{S}=const$ - the dipole picture 
 suggested in \cite{M1}. 

{\bf 3. Perturbative and non-perturbative.}

The perturbative gluons are confined and 
do not propagate to large distances. Available fits \cite{MEGGIO} to the lattice QCD 
data suggest Yukawa screening of perturbative color fields with 
propagation/screening 
radius $R_c\approx 0.2-0.3$ fm. The value $R_c=0.275$ fm has been used since 1994 
in the very successful color dipole phenomenology of small-x DIS 
\cite{NZHERA,NZZEXP,NZZ97,SlopeJETP98,CharmBeauty}. 
Because the propagation radius is short compared to the typical range of strong 
interactions the dipole cross section obtained as a solution of the CD BFKL
 equation
(\ref{eq:CDBFKL}) would miss the interaction strength for large color dipoles. 
In \cite{NZHERA,NZZEXP} this missing strength was modeled by the $x$-independent
 dipole cross section and it has been assumed that the perturbative, 
$\sigma(\xi,r)$,  and non-perturbative, $\sigma_{npt}(r)$, cross sections 
are additive,
\beq
\sigma_{tot}(\xi,r)=\sigma(\xi,r)+\sigma_{npt}(r).
\label{eq:PTNPT}
\eeq
\begin{figure}[h]
\psfig{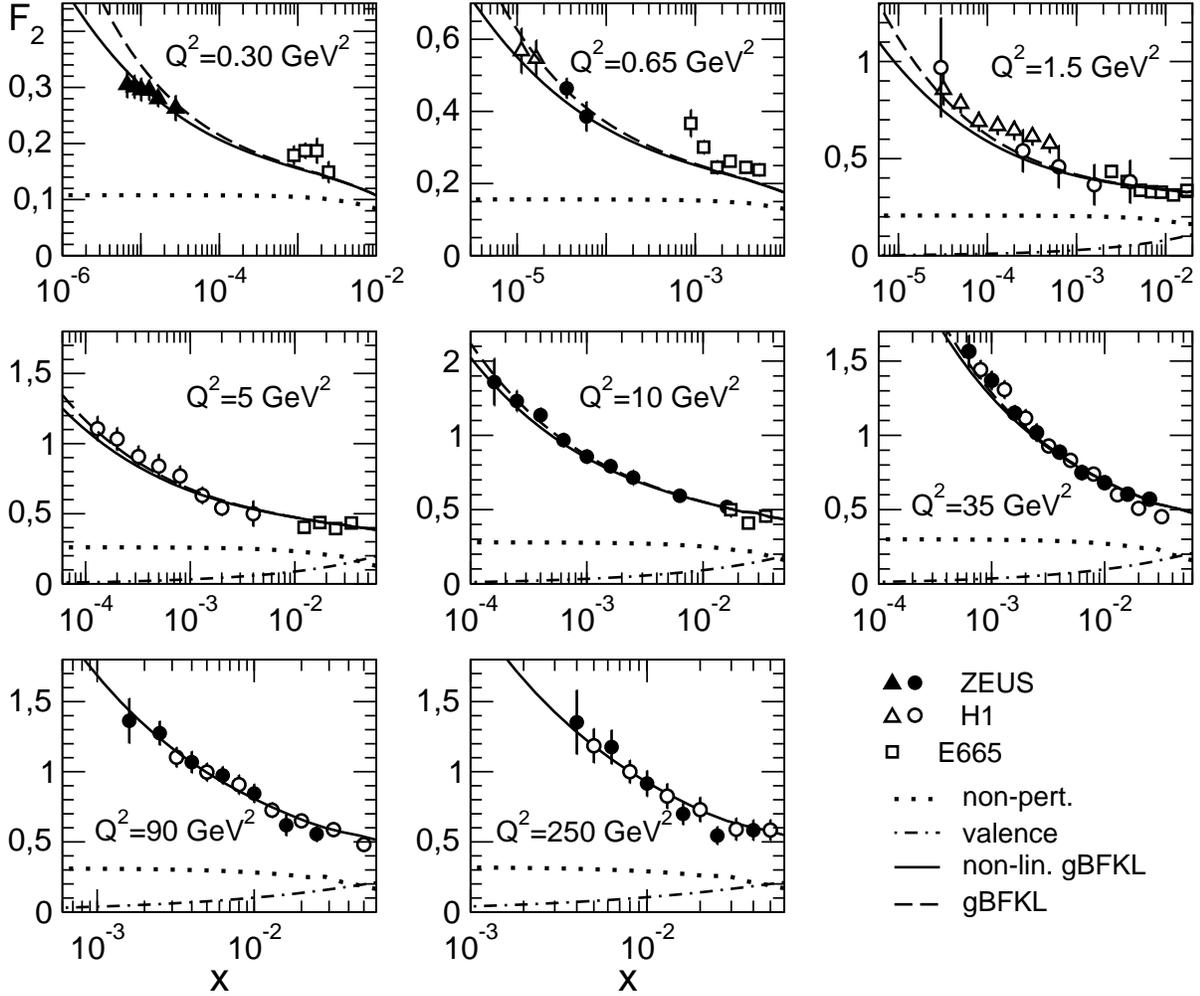}
\vspace{-0.5cm}
\caption{The CD BFKL description 
 of the experimental data on $F_2(x,Q^2)$.  
 Black triangles and circles are ZEUS data 
\cite{ZBREIT97,ZDER96}, open triangles  and circles show H1 data
 \cite{H1ADL96,H1AID96} and open 
squares refer to E665 results \cite{E665}. 
Dashed lines represent the linear CD BFKL structure function $F_2$.
Shown by  solid lines are the  non-linear CD BFKL structure functions 
 $F_2$. At high $Q^2$ the non-linear effects vanish and 
 both dashed and solid lines are indistinguishable.
 The valence and non-perturbative corrections are 
included into both the CD BFKL and the non-linear CD BFKL description of $F_2$. 
The contribution to $F_2$ from DIS off valence
 quarks \cite{GRV98} is shown separately  by dash-dotted lines. 
Shown by  dotted lines are 
the non-perturbative contributions to $F_2$.} 
\label{fig:fig1}
\end{figure}  
The principal point about the non-perturbative component of $\sigma_{tot}(\xi,r)$
is that it must not be subjected to the perturbative BFKL evolution. Thus, the 
arguments about the rise of $\sigma(\xi,r)$ due to the hard-to-soft diffusion
 do not apply to  $\sigma_{npt}(r)$. 
We reiterate, finite $R_c$ means that gluons with
the wave length $\lambda \gsim R_c$ are beyond  the realm of  
perturbative QCD. A quite common application  of purely perturbative 
non-linear equations \cite{B,K} to the analysis of DIS data without proper 
separation of perturbative and non-perturbative contributions is 
completely unwarranted.

Specific form of $\sigma_{npt}(r)$ motivated by the QCD string 
picture and used in the present paper is as follows:
\beq
\sigma_{npt}(r)=a\alpha^2_S(r)r^2/(r+d).
\label{eq:NPT}
\eeq
Here $d=0.5$ fm  is close to the radius
of freezing of the running QCD coupling $r_f$  and $a=5.$ fm.

Our choice $R_{c}=0.26$\,fm  leads to a very good
description  of the  data \cite{ZBREIT97,ZDER96,H1ADL96,H1AID96,E665} 
 on the proton structure
function  $F_2(x,Q^2)$ at small $x$ shown in Fig.\ref{fig:fig1}.
Shown separately are the nonperturbative contribution (\ref{eq:NPT})
and the contribution  from DIS off valence
 quarks \cite{GRV98}.
The effects of quark masses important at low $Q^2$ are taken into 
account \cite{FZJL2011}. The linear CD BFKL description of $F_2(x,Q^2)$ 
(dashed line) is perfect at moderate and high $Q^2$ where it is 
indistinguishable from the solid
line representing the non-linear CD BFKL results (see below).
Two lines diverge at low $Q^2$ where the account of the  
non-linear effects    improves the agreement with data.

Recently a global analysis of HERA DIS data has been reported 
\cite{AAA}. In \cite{AAA} a purely perturbative non-linear equation
 is solved with  some phenomenological initial conditions. 
A very soft infrared regularization with the infrared cutoff 
$\sim \Lambda^{-1}_{QCD}$  allows 
non-perturbatively large dipoles to be  governed by 
the perturbative QCD dynamics. The non-perturbative component 
of solution evolves perturbatively to smaller $x$.
 Good agreement with data was found.

{\bf 4. CD BFKL and the partial-wave amplitudes.}

Following  \cite{SL94,SLPL} we 
rewrite  the  Eq.(\ref{eq:CDBFKL}) in terms of 
 the $q\bar q$-nucleon  partial-wave amplitudes 
(profile functions)  $\Gamma(\xi,\br,\bb)=1-S(\xi,\br,\bb)$ related to the 
scattering matrix $S(\xi,\br,\bb)$. We 
introduce the impact parameter ${\bb}$ defined with respect
to the center of the $q$-$\bar{q}$ dipole. In the $q\bar{q}g$
state, the $qg$ and $\bar{q}g$ dipoles
have the impact parameter ${\bf b}+{\brho}_{2,1}/2$. In the large-$N_c$ 
approximation  
$\sigma_3$ in
Eq.~(\ref{eq:SIGMA3}) reduces to 
$\sigma_3= \sigma(\xi,\brho_1)+\sigma(\xi,\brho_2)$.
what corresponds to  the factorization of the 3-parton ($q\bar q g$) 
scattering  matrix,
\beq
S_3(\xi,\br,\brho_1,\brho_2)=S(\xi,\rho_{1},{\bf b}+{1\over 2}{\brho}_{2})
S(\xi,\rho_{2},{\bf b}+{1\over 2}{\brho}_{1}).
\eeq 
Then, the renormalization of the $q\bar q$-nucleon scattering matrix, 
$S(\xi,\br,\bb)$, 
for the perturbative $3$-parton state $q\bar q g$ is as follows
\bea
{\partial_{ \xi} S(\xi,r,\bb)}
=\int d^{2}{\brho}_{1}\,\,
\left|\psi({\brho}_{1})-\psi({\brho}_{2})\right|^{2}\nonumber\\
\times\left[S(\xi,\rho_{1},{\bf b}+{1\over 2}{\brho}_{2})
S(\xi,\rho_{2},{\bf b}+{1\over 2}{\brho}_{1})-S(\xi,r,\bb)\right].
\label{eq:SBFKL}
\eea
For the early  discussion of  Eq.~(\ref{eq:SBFKL}) see \cite{B,K}.
The substitution $S(\xi,\br,{\bf b})=1-\Gamma(\xi,\br,{\bf b})$ results in
\bea
{\partial_{ \xi} \Gamma(\xi,r,{\bf b})}
= \int d^{2}{\brho}_{1}
\left|\psi({\brho}_{1})-\psi({\brho}_{2})\right|^{2}\nonumber\\
\times\left[\Gamma(\xi,\rho_{1},{\bf b}+{1\over 2}{\brho}_{2}) +
\Gamma(\xi,\rho_{2},{\bf b}+{1\over 2}{\brho}_{1}) -
\Gamma(\xi,r,{\bf b})\right]\nonumber\\
-\Gamma(\xi,\rho_{2},{\bf b}+{1\over 2}{\brho}_{1})
\Gamma(\xi,\rho_{1},{\bf b}+{1\over 2}{\brho}_{2}).
\label{eq:GAMBFKL}
\eea 
We identify the corresponding partial waves using the conventional
impact parameter representation  for the 
 elastic dipole-nucleon amplitude 
\beq
f(\xi,r,\bk)=2\int d^2{\bf b}\exp(-i\bb\bk) \Gamma(\xi,r,{\bf b}).
\label{eq:AMPL}
\eeq
For the predominantly imaginary $f(\xi,r,\bk)=i\sigma(\xi,r)\exp(-Bk^2/2)$
the profile function is
\beq
\Gamma(\xi,r,{\bf b})={\sigma(\xi,r)\over 4\pi B(\xi,r)}
\exp\left[-{b^2\over 2B(\xi,r)}\right].
\label{eq:GAMMA}
\eeq
and $\sigma(\xi,r)=2\int d^2{\bf b}\, \Gamma(\xi,r,{\bf b})$.

Integrating over ${\bf b}$  Eq. (\ref{eq:GAMBFKL}) 
yields \cite{FZSLOPE}
\bea
{\partial_{\xi}\sigma(\xi,r)} =
 \int d^{2}{\brho}_{1}\,\,
\left|\psi({\brho}_{1})-\psi({\brho}_{2})\right|^{2}\nonumber\\
\times\left\{\sigma(\xi,\rho_{1})+
\sigma(\xi,\rho_{2})-\sigma(\xi,r)\right.\nonumber\\
\left.-{\sigma(\xi,\rho_{1})\sigma(\xi,\rho_{2})\over 4\pi(B_1+B_2)}
\exp\left[-{r^2\over 8(B_1+B_2)}\right]\right\},
\label{eq:BFKLNL}
\eea
where $B_i=B(\xi,\rho_i)$.
The above  definition of the scattering profile function, Eq.~(\ref{eq:GAMMA}),
  removes uncertainties with the radius $R$ of the area within which 
 interacting gluons  are expected to be distributed (the parameter 
$S_{\perp}=\pi R^2$
appearing in Eq. (\ref{eq:GLR2})). In different analyses of 
the non-linear effects its value varies from the realistic $R^2=16$ GeV$^{-2}$ 
\cite{KK2003} down to the intriguing small $R^2=3.1 $ GeV$^{-2}$  \cite{Bartels}.
Besides, the radius $R$ is usually assumed to be independent of $x$.
In our approach the area populated with interacting gluons is 
proportional to the diffraction cone slope $B(\xi,r)$.

{\bf{5. The diffraction cone slope.}}

The diffraction slope 
 for the forward cone in the dipole-nucleon scattering \cite{SL94}
was presented in \cite{SLPL} in a very symmetric form  
\beq 
B(\xi,r)={1\over 2}\langle {\bf b}^2\rangle ={1\over 8}r^2+{1\over 3} R_N^2 
+2\alpha^{\prime}_{\Pom}\xi.
\label{eq:BSLOPE}
\eeq
The latter  provides 
the  beam, target and exchange decomposition of $B$:
 $r^2/8$ is the purely geometrical term for  the 
color dipole of the size $r$, $R_N$ represents
the gluon-probed radius of the proton, 
the dynamical 
component of $B$ is given by the last term in Eq. (\ref{eq:BSLOPE}) 
 where 
$\alpha^{\prime}_{\Pom}$ is the Pomeron trajectory slope evaluated first 
in \cite{SL94} (see also \cite{SLPL}). 
The order of magnitude estimate \cite{SLPL} 
\beq
\alpha^{\prime}_{\Pom}\sim
{3 \over 16\pi^{2}} \int d^{2}\vec{r}\,\,\alpha_{S}(r)
R_c^{-2}r^{2}
K_{1}^{2}(r/R_c)  \sim
{3 \over 16\pi}\alpha_{S}(R_{c})R_{c}^{2}
\, ,
\label{eq:ALPRIME}
\endeq
 clearly shows the connection between the dimensionful
$\alpha^{\prime}_{\Pom}$ and the non-perturbative infrared parameter
$R_{c}$. 
The  increase of $B$ with growing collision energy is  
known as the phenomenon of shrinkage of the diffraction cone. 

We determine $\alpha^{\prime}_{\Pom}$ 
as  the   $\xi \rightarrow \infty$ limit of the local Regge slope 
$\alpha_{eff}^{\prime}(\xi,r)=
{\partial_{ \xi} B(\xi,r)/2}$ \cite{SLPL}.
At $\xi \rightarrow \infty$,
$\alpha^{\prime}_{eff}(\xi,r)$ tends to a $r$-independent
$\alpha^{\prime}_{\Pom}=0.064$ GeV$^{-2}$.
The  onset of the limiting value
$\alpha^{\prime}_{\Pom}$ is very slow and
correlates nicely with the very slow onset
of the BFKL asymptotics of $\sigma(\xi,r)$ \cite{NZZJL94}.
An interesting finding of Ref. \cite{SLPL} is a large sub-asymptotic 
value of the
effective Regge slope $\alpha^{\prime}_{eff}(\xi,r)$,
which is by the factor $\sim(2-3)$ larger than
$\alpha^{\prime}_{\Pom}$. 

In Eq. (\ref{eq:BSLOPE}) the gluon-probed radius of the proton is a 
phenomenological parameter to be determined from the experiment.
The analysis of Ref. \cite{INS2006} gives
$R_N^2 \approx 12{\, \rm GeV}^{-2}$.

\begin{figure}[h]
\psfig{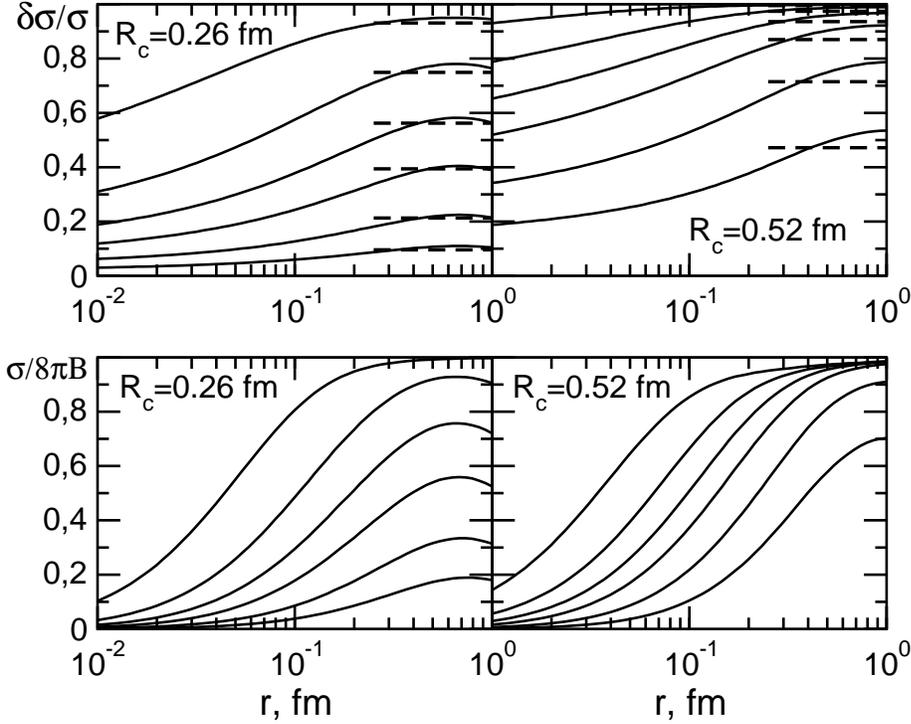}
\vspace{-0.5cm}
\caption{The dipole size dependence of the non-linear correction 
$\kappa=\delta\sigma/\sigma$ to
the linear CD BFKL dipole cross section $\sigma$ for two correlation radii
$R_c$ and for $\xi=6$, $8.5$, $11$, $13$, $15.5$, $20$.
Dashed lines correspond to the approximation 
$v(\xi)\approx\sigma_0(R_c^2)(e^{c\xi}-1)$ in 
Eqs.~(\ref{eq:LARGE3},\ref{eq:LARGE4}).
Shown separately is the  ``unitarity ratio'' 
$\sigma/8\pi B$ (see text)
for two values of $R_c$ and for the same set of  $\xi$. }
\label{fig:fig2}
\end{figure}

{\bf 6.  Non-linear CD BFKL: small dipoles, $r\ll R_c$.} 

The term quadratic in $\sigma$ in  Eq.~(\ref{eq:BFKLNL}),  
models the process of the gluon fusion.
The efficiency of this ``fuser'' 
differs substantially  for $r\ll R_c$ and for   $r\gsim R_c$.
Consider first the ordering of dipole sizes
\beq
r^2\ll\rho^2\ll R_c^2
\label{eq:ORDER}
\eeq
corresponding to the Double Leading Log Approximation (DLLA) \cite{DGLAP}.
  Eq.(\ref{eq:BFKLNL}) reduces to
\bea
{\partial_{\xi} \sigma(\xi,r)} =
{C_F\over \pi}\alpha_S(r)r^2\times\nonumber\\
\times\int_{r^2}^{R_c^2} {d\rho^2\over \rho^4}   
\left[2\sigma(\xi,\rho)-{\sigma(\xi,\rho)^2\over 8\pi B}\right].
\label{eq:DLLANL}
\eea
 First notice that the function  
\beq
g(\xi,\rho)=\rho^{-2}\sigma(\xi,\rho)
\label{eq:gsoft}
\eeq
is essentially flat in $\rho$ and the second term  in the  $rhs$ of
Eq. (\ref{eq:DLLANL}) is dominated by 
$\rho\sim R_c$,
\bea
{1\over 8 B}\int_{r^2}^{R_c^2} {d\rho^2\over \rho^4}
\sigma(\xi,\rho)^2
 \simeq
{R_c^2\over 8B}g(\xi,R_c)^2
\label{eq:G2INT}
\eea
with $B=B(\xi,R_c)$. Thus, the new  dimensionless parameter
\beq
\kappa_c={R_c^2\over 8B}
\label{eq:GEOM} 
\eeq
enters the game. Its geometrical meaning is quite clear. Remind
 that the unitarity requires (see Eq.~(\ref{eq:GAMMA})) 
\beq
\sigma(\xi,\rho)\leq 8\pi B.
\label{eq:UNIT}
\eeq
 Smallness of $\kappa_c$  
 makes the non-linear effects rather weak at HERA
even at lowest available Bjorken $x$ (see Fig.~\ref{fig:fig1}). 
Comparison of the linear and quadratic terms in the right hand side of 
Eq.~(\ref{eq:DLLANL})
shows that the relative strength  of non-linear effects 
decreases  to smaller $r^2\sim Q^{-2}$  logarithmically
\beq
\kappa={quadr.\over lin.}
\propto {\kappa_c\ln^{-1} (Q^2R_c^2)},
\label{eq:KAPPALOG}
\eeq
 Therefore, we are dealing with the scaling rather than 
the higher twist, $1/Q^2$, effect.

{\bf 7. Saturation scale and observables.} 
 
The parameter $\kappa$ in Eq.~(\ref{eq:KAPPALOG}) should not be confused 
with another parameter frequently  used  to quantify 
the strength of the non-linear effects.
It decreases with growing $Q^2$ much faster than $\kappa$ in 
Eq.~(\ref{eq:KAPPALOG}). Namely, 
\beq
\kappa\sim {1\over Q^2}.
\label{eq:KAPPA2}
\eeq
The estimate (\ref{eq:KAPPA2}) comes from equating the linear and non-linear 
terms in $rhs$ of the equation \cite{GLR,MQ}
\beq
{\partial_{\xi}\partial_{\eta}G(\xi,\eta)} = 
c G(\xi,\eta)-{a\over Q^2}G(\xi,\eta)^2,
\label{eq:GLR2}
\eeq
where $\eta=\ln(Q^2/\Lambda_{QCD}^2)$, 
$c=\alpha_S N_c/\pi$, 
$a=\alpha_S^2\pi/S_{\perp}$, $G(\xi,\eta)$ is the integrated gluon density 
and Eq.(\ref{eq:GLR2}) comes from Eq.(\ref{eq:DLLANL}) as  for small dipoles
$g(\xi,\rho)\approx {\pi^2\over N_c}\alpha_S(\rho)G(\xi,\rho)$.
The corresponding value of $Q^2$ denoted by
\beq
Q^2_s=aG(x,Q^2_s)/c
\label{eq:SATSCALE}
\eeq
 is called the saturation scale. 
The non-linear saturation effects are assumed to be substantial for all 
$Q\lsim Q_s$ (see e.g. \cite{JK}). 
Obviously,  Eq.~(\ref{eq:KAPPALOG}) asserts something different. 
The point is that 
Eqs.~(\ref{eq:KAPPALOG}) and (\ref{eq:KAPPA2}), 
describe the $Q^2$-dependence of  strength of the non-linear effects
for  two very different quantities: the integrated 
gluon density $G(\xi,\eta)$ and the differential gluon density 
${\cal F}(\xi,\eta)=\partial_{\eta}G(\xi,\eta)$, respectively. 
The gluon density  
$G(\xi,\eta)$ is 
a directly measurable quantity. For example, the longitudinal DIS 
structure function is  
$F_L(x,Q^2)\sim \alpha_S(Q^2)G(x,Q^2)$ \cite{FLDOK}. On the contrary, 
the differential gluon density ${\cal F}(\xi,\eta)$ is related to the observable 
quantities like $F_2(x,Q^2)$ rather indirectly, 
by means of the well known transformations  leaving  a  weak trace of 
Eq.~(\ref{eq:KAPPA2}) in $F_2(x,Q^2)$.

A possibility to test Eqs.~(\ref{eq:KAPPA2}) and (\ref{eq:SATSCALE})
provides the coherent diffractive dijet production in pion-nucleon and 
pion-nucleus collisions \cite{NSS2000}. Both helicity amplitudes  
of the process are directly proportional to ${\cal F}(x,k^2)$. 
The same proportionality of diffractive amplitudes to ${\cal F}(x,k^2)$
was found for real photoproduction with pointlike $\gamma q\bar q$ vertex 
in \cite{NZDIJET}. Therefore, there is no real clash between 
Eqs.~(\ref{eq:KAPPALOG}) and (\ref{eq:KAPPA2}). 
The sharp $Q^2$-dependence of the nonlinear term in Eq.~(\ref{eq:GLR2})
does not imply vanishing non-linear effects in $G(x,Q^2)$ for $Q^2\gg Q^2_s$.

{\bf 8. Non-linear CD BFKL: large dipoles, $r\gsim R_c$.}
 
The interplay of the color screening and gluon fusion effects 
 at large  $r\gsim R_c$, where the non-linear effects are expected to be
 most pronounced, requires special investigation. In high-energy scattering
 of  large quark-antiquark dipoles, $r\gg R_c$,
 a sort of the additive quark model is recovered:
the (anti)quark of the dipole $r$ develops its own perturbative 
gluonic cloud and the pattern of diffusion changes dramatically. Indeed,
in this region the term proportional to 
$K_1(\rho_1/R_c)K_1(\rho_2/R_c)$ in the 
kernel of Eq. (\ref{eq:CDBFKL}) is exponentially small, what is related 
to the exponential decay
of the correlation function (the propagator) of perturbative gluons. 
Then, at large $r$ the kernel  will be dominated by the 
contributions from $\rho_1\lsim R_c\ll\rho_2\simeq r$ and from
$\rho_2\lsim R_c\ll\rho_1\simeq r$. It does not depend on $r$ and 
 for large $N_c$ the equation for the dipole cross section reads
\bea
{\partial_{\xi}\sigma(\xi,r)}={\alpha_S C_F\over \pi^2}\int
 d^2\rho_1  R_c^{-2}K_1^2(\rho_1/R_c)\nonumber\\
\left\{\sigma(\xi,\rho_1)+\sigma(\xi,\rho_2)
-\sigma(\xi,r)\right.\nonumber\\
\left.-{\sigma(\xi,\rho_1)\sigma(\xi,\rho_2)\over 4\pi(B_1+B_2)}
\exp\left[-{r^2\over 8(B_1+B_2)}\right]\right\},
\label{eq:LARGE1}
\eea 
where $B_i=B(\xi,\rho_i)$.  
For a qualitative understanding of the role of color screening 
in the non-linear dynamics of large dipoles
 we  reduce Eq.~(\ref{eq:LARGE1}) to the 
differential equation.
  First notice that the dipole cross section $\sigma(\xi,r)$ as a function 
of $r$ varies slowly in the  region $r\gg R_c$, 
while the function $K_1(y)$ vanishes exponentially at $y\gg 1$ and
  $K_1(y)\approx 1/y$ for $y\ll 1$.  
Therefore, Eq.~(\ref{eq:LARGE1}) can be cast in the following form
\bea
c^{-1}{\partial_{\xi}\sigma(\xi,r)}=\sigma(\xi,R_c)
+R_c^2{\partial_{r^2}\sigma(\xi,r)}-\nonumber\\
-\sigma(\xi,R_c)\sigma(\xi,r)/8\pi B,
\label{eq:LARGE2}
\eea
where $c=\alpha_SC_F/\pi$ and for simplicity $B=B(0,R_c)$.
The solution of Eq.~(\ref{eq:LARGE2})  with the boundary condition
 $\sigma(0,r)=\sigma_0(r^2)$, where $\sigma_0(r^2)$ comes from 
Eq. (\ref{eq:BORN}), is 
\beq
\sigma(\xi,r)=
{\sigma_0(r^2+c\xi R_c^2)+v(\xi)
\over 1+v(\xi)/8\pi B},
\label{eq:LARGE3}
\eeq
where
\beq
v(\xi)=e^{c\xi}\int^{c\xi}_0\sigma_0(R_c^2+zR_c^2)e^{-z}dz.
\label{eq:LARGE4}
\eeq

From Eq.~(\ref{eq:BORN}) it follows that 
at $r\ll l= \min\{R_c/\sqrt{2},\sqrt{u}/\Lambda\}$ 
\beq
\sigma_{0}(r^2)\approx {4\pi^2C_F\over \beta_0}r^2\alpha_S(r/\sqrt{A})
\ln{\alpha_S(l)\over \alpha_S(r/\sqrt{A})},
\label{eq:BORN1}
\eeq
where  $A\approx 10$ comes from properties of the Bessel function $J_0(y)$ in 
Eq.~(\ref{eq:BORN}) \cite{NZA10}.
 For large dipoles $\sigma_{0}(r^2)$ saturates at
$r^2\simeq A l^2$, 
\beq
\sigma_{0}(r^2)\approx 4\pi C_F R_c^2\alpha_f\alpha_S(R_c) h(a),
\label{eq:BORN2}
\eeq
where
$\alpha_f=0.8$ (see Eq.(\ref{eq:ALPHAS})) and 
 the interplay of two scales, $R_c$ and $\Lambda^{-1}$, in 
Eq.~(\ref{eq:BORN}) results in 
 $h(a)=1-(a^2-1-2a\ln{a})/(a-1)^3$ with $a=u/R_c^2\Lambda^2$. 
This kind of saturation is due to the finite
propagation radius of perturbative gluons.

With growing $\xi$ the dipole cross section $\sigma(\xi,r)$  
increases approaching the unitarity bound, $\sigma=8\pi B$.
To  quantify  the strength of the non-linear effects we  introduce
the parameter
\beq
\kappa={\delta\sigma/\sigma},
\label{eq:kappa}
\eeq
where $\delta\sigma=\sigma-\sigma_{nl}$ and 
$\sigma$ represents the solution of the linear CD BFKL 
Eq. (\ref{eq:CDBFKL}), while 
 $\sigma_{nl}$ stands for the solution of the non-linear CD BFKL
 Eq.(\ref{eq:BFKLNL}). 
Therefore, our $\kappa$ gives the strength 
of the non-linear effects with the non-perturbative corrections 
switched off
\bea
\kappa= {\delta\sigma/\sigma}={v(\xi)\over {v(\xi)+8\pi B}}\nonumber\\
\sim {4\pi\kappa_c}{C_F\over\beta_0}\alpha_S(R_c/\sqrt{A})(e^{c\xi}-1).
\label{eq:kappa2}
\eea
The magnitude of 
non-linear effects 
is  controlled, like in the case of small dipoles, by  the ratio $R_c^2/B$
(we assumed $R_c^2\ll u/\Lambda^2$).

Numerical solution of Eqs.(\ref{eq:CDBFKL},\ref{eq:BFKLNL}) gives
 the $r$-dependence of $\kappa={\delta\sigma/\sigma}$ shown in
 Fig.~\ref{fig:fig2} for several
values of $x$ and for two correlation radii  
$R_c=0.26$ fm and $R_c=0.52$ fm. For
${\delta\sigma/\sigma}\ll 1$  the law  
${\delta\sigma/\sigma}\propto R_c^2/B$ holds true. 
At large $r\gsim R_c$ the toy-model solution, 
Eq.~(\ref{eq:LARGE3}), (dashed lines) correctly reproduces the 
$\xi$-dependence of $\kappa$. 
At small $r\ll R_c$ the ratio ${\delta\sigma/\sigma}$ decreases slowly as 
it is prescribed by 
Eq.~(\ref{eq:KAPPALOG}). 
In Fig.~\ref{fig:fig2} also shown is  the  evolution of 
 the unitarity ratio $\sigma_{nl}(\xi,r)/(4\pi(B_1+B_2))$ with 
$B_1=B(\xi,R_c)$, $B_2=B(\xi,r)$ and
denoted  by $\sigma/8\pi B$. 
High sensitivity of
$\sigma_{nl}(\xi,r)$ to $R_c$ is not surprising in view of
the toy-model
solution (\ref{eq:LARGE3}).

{\bf 9. Summary.}
 
To summarize, the purpose of the present paper has been an
exploration of the phenomenology of saturation in  diffractive 
scattering which emerges from the BFKL dynamics with finite correlation
 length of perturbative gluons, $R_c$. The non-linear effects are
 shown to be dominated by the large size $q\bar q-g$ fluctuations
of the probe (virtual gauge boson). They should very slowly, 
$\propto 1/\ln Q^2$, decrease at large $Q^2$.
The magnitude of 
the non-linear effects is controlled by the dimensionless parameter
 $\kappa_c=R_c^2/8B$. The area 
populated with interacting gluons is 
proportional to the diffraction cone slope $B$. Smallness of $\kappa_c$  
 makes the non-linear effects rather weak 
even at lowest  Bjorken $x$ available at HERA.
The linear BFKL with the running coupling and the infrared 
regulator $R_c=0.26$ fermi gives very good description of the 
proton structure function $F_2(x,Q^2)$ in a wide range 
of $x$ and $Q^2$.

{\bf Acknowledgments.} 
V.R.~Z. thanks N.N. Nikolaev and 
B.G. Zakharov for useful discussions and 
 the Dipartimento di Fisica dell'Universit\`a
della Calabria and the INFN - 
gruppo collegato di Cosenza for their warm
hospitality while a part of this work was done.
The work was supported in part by the Ministero Italiano
dell'Istruzione, dell'Universit\`a e della Ricerca,   by
 the RFBR grants 11-02-00441, 12-02-00193 and by the 
DFG grant 436 RUS 113/940/0-1.

\end{document}